\documentclass[journal=jacsat,manuscript=article]{achemso}

\usepackage[T1]{fontenc}       % Use modern vector font encodings (not bmp characters)
\usepackage{lmodern} % use latin modern fonts 
\usepackage{amsmath, amsthm, amscd, amssymb} % the AMS packages
\usepackage[utf8]{inputenc} % most of not english characters!
\usepackage[T1]{fontenc} % accents and so on
\usepackage{graphicx} % for importing pictures
\usepackage{hyperref} % turn all your internal references into hyperlinks
\usepackage{natbib} % Journal style bibliography management. Bibliography files in the bibtex syntax are expected
\usepackage{achemso}

\usepackage{xcolor}          %%%
\usepackage[normalem]{ulem}  %%%
\author{Gerard P. Conangla}
\email{gerard.planes@icfo.eu}
\author{Andreas W. Schell}
\author{Raúl A. Rica}
\altaffiliation{Department of Applied Physics, School of Sciences, University of Granada, 18071, Granada, Spain
}
\author{Romain Quidant}
\email{romain.quidant@icfo.eu}
\affiliation{ICFO-Institut de Ciencies Fotoniques, The Barcelona Institute of Science and Technology, 08860 Castelldefels, Barcelona, Spain}
\alsoaffiliation{ICREA-Institució Catalana de Recerca i Estudis Avançats, 08010 Barcelona, Spain}

\title{Motion control and optical interrogation of a levitating single NV in vacuum}

\abbreviations{}

\begin{document}
% \maketitle is automatic

%%%%%%%%%%%%%%%%%%%%%%%%%%%%%%%%%%%%%%%%%%%%%%%%%%%%%%%%%%%%%%%%%%%%%
%% ABSTRACT
%%%%%%%%%%%%%%%%%%%%%%%%%%%%%%%%%%%%%%%%%%%%%%%%%%%%%%%%%%%%%%%%%%%%%

\begin{abstract} %% 7th
Levitation optomechanics exploits the unique mechanical properties of trapped nano-objects in vacuum in order to address some of the limitations of clamped nanomechanical resonators. In particular, its performance is foreseen to contribute to a better understanding of quantum decoherence at the mesoscopic scale as well as to lead to novel ultra-sensitive sensing schemes. While most efforts have so far focused on optical trapping of low absorbing silica particles, further opportunities arise from levitating objects with internal degrees of freedom like color centers. Nevertheless, inefficient heat dissipation at low pressures poses a challenge, as most nano-objects, even with low absorbing materials, experience photo-damage in an optical trap. Here, by using a Paul trap, we demonstrate levitation in vacuum and center-of-mass feedback cooling of a nanodiamond hosting a single nitrogen-vacancy center. The achieved level of motion control enables us to optically interrogate and characterize the emitter response. The developed platform is applicable to a wide range of other nano-objects and represents a promising step towards coupling internal and external degrees of freedom.
\end{abstract}

%%%%%%%%%%%%%%%%%%%%%%%%%%%%%%%%%%%%%%%%%%%%%%%%%%%%%%%%%%%%%%%%%%%%%
%%\section{Introduction} 4th
%%%%%%%%%%%%%%%%%%%%%%%%%%%%%%%%%%%%%%%%%%%%%%%%%%%%%%%%%%%%%%%%%%%%%

% general goal

Optomechanics offers a toolbox to investigate classical and quantum mechanical oscillators in a highly controlled way. A fundamental open question in the field is under which conditions the transition from quantum to classical behaviour takes place, but despite the high level of control achieved with micro and nanomechanical systems, a good understanding of this transition remains elusive. Experiments so far have demonstrated the preparation of mechanical oscillators close to their ground state\citep{Oconnell2010}\citep{Chan2011}\citep{Teufel2011}\citep{Aspelmeyer2014}, but the controlled generation of arbitrary quantum motional states\citep{Armour2002} greatly increases the technical requirements. An exciting route towards these ends is the use of nanoscale mechanical resonators containing internal degrees of freedom\citep{Rabl2009}\citep{Teissier2014}\citep{lee2017}, whose energy levels can be coupled to the motion of the oscillator. This scheme would allow state transfer or cooling and outperform classical resonators, for instance in ultra-sensitive sensing\citep{Rabl2009}. 

% Levitated particles as resonators, and problems with optical traps

In this regard, nanomechanical resonators based on nanoparticles levitated in vacuum are especially attractive, because they are highly decoupled from the environment due to the absence of clamping, and hence exhibit very large quality ($Q$) factors, even at moderate pressures. In optical traps, optical forces can be used to efficiently cool the nanoparticle center-of-mass (COM) motion and thus reduce the influence of thermal noise\citep{Li2011}\citep{Gieseler2012}. Furthermore, one could add further functionality and control to the platform by levitating functional nano-objects with tailored specific properties. For instance, levitated particles with internal degrees of freedom (DOF), such as controllable spin systems, have the potential to be used in matter-wave interferometry\citep{Scala2013}\citep{Albrecht2014}\citep{Yin2013}. However, the optical levitation of nanoparticles with internal DOF faces some challenges\citep{Neukirch2015}\citep{Rahman2016}: on one hand, large optical trapping powers introduce a big constraint on the type of particles that can be levitated in vacuum, since heat from residual absorption of the trapping laser can not be efficiently dissipated at low pressures. On the other hand, high vacuum levels are necessary for maintaining motion coherence and attaining large $Q$ factors. Hence, to avoid photo-damage other methods to levitate particles with internal DOF are needed.

% why use Paul traps

An alternative approach to optical trapping is the use of Paul traps\citep{Paul1990}, which have been widely applied to manipulate individual ions. The main appeal of Paul traps is the possibility to levitate charged particles without the aid of optical fields \citep{Nagornykh2015}\citep{Millen2015}\citep{Alda2016}, thereby widening the range of their constitutive materials. Paul traps are thus well suited for the levitation of particles with optical defects, like nanodiamonds hosting Nitrogen-Vacancy (NV) centers. NV centers are optical emitters formed by a nitrogen impurity and a vacancy in the crystal lattice of diamond\citep{Aharonovich2016}. NV centers are attractive because they are a stable source of single photons even at room-temperature, whose electron spin can be optically addressed and possesses long coherence times\citep{Bar2013}. They have already been successfully used as a qubit\citep{Togan2010}\citep{Pfaff2014}\citep{Hensen2015} and are excellent candidates for sensing electric and magnetic fields by either using Stark or Zeeman shifts\citep{Taylor2008}\citep{Hall2009}\citep{Dolde2011}. So far, Paul traps have been used to trap and optically interrogate small nanodiamonds clusters at ambient pressures\citep{Kuhlicke2014} and larger microdiamonds\citep{Delord2017} hosting multiple nitrogen vacancy (NV) centers. Nevertheless, levitation and detection of a single NV center in vacuum, in a strongly underdamped regime of the nanoparticle's oscillation, has not yet been achieved, requiring a further level of control on the particle dynamics compatible with a single photon optical detection.

% what we do and why it is so cool

\begin{figure}
\includegraphics[width=0.8\textwidth]{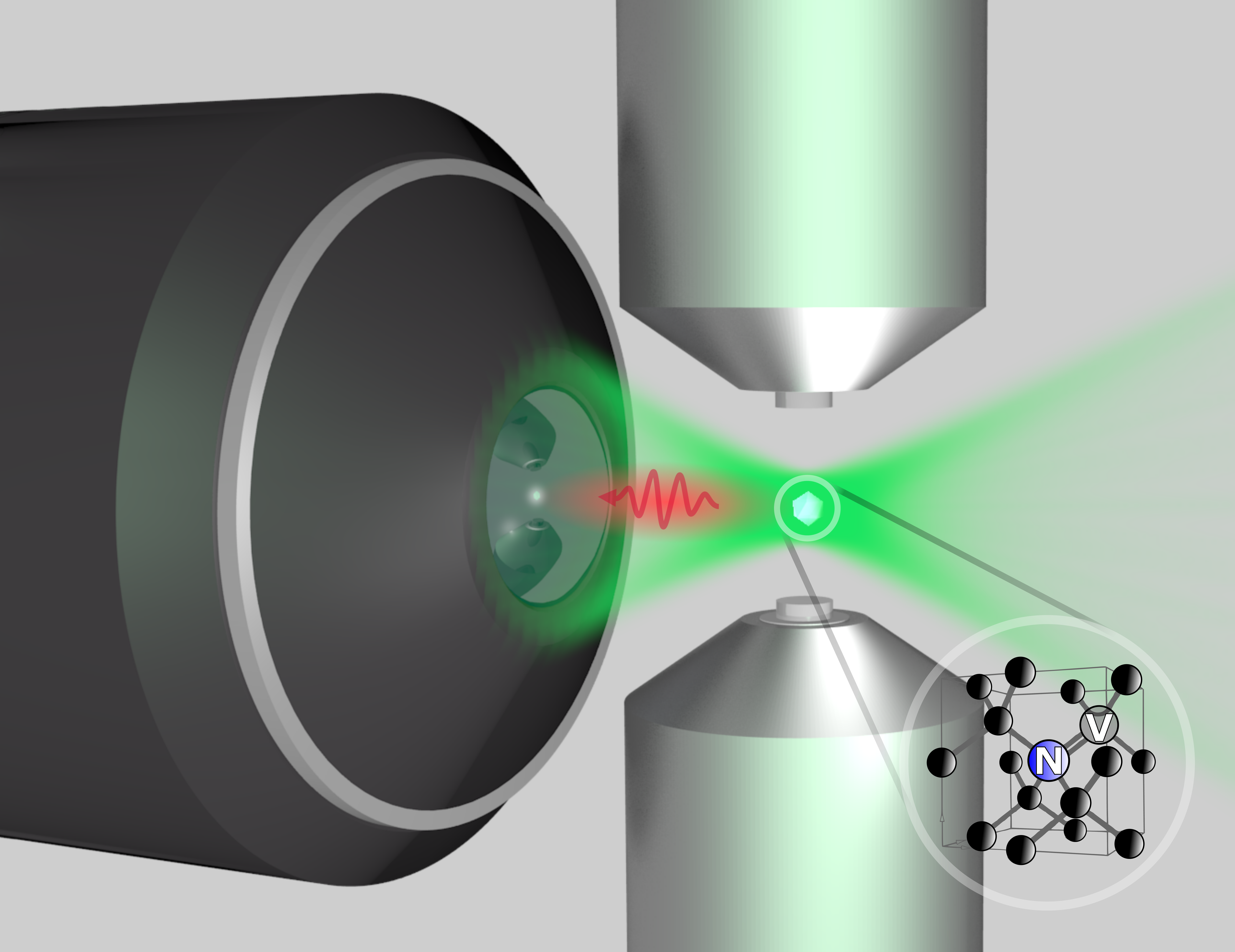}
\caption{Artistic representation of the trapping scheme. A charged nanodiamond containing a single 
NV center sits at the minimum of the effective potential of a Paul trap. The nanodiamond is illuminated 
with a continuous wave 532 nm laser with a high numerical aperture objective lens. The fluorescence emitted 
by the NV center is collected using the same objective lens. The simplified atomic structure of the NV center is also shown.}
\label{fig:paultrap}
\end{figure}

In this letter we report on two key requirements for coupling internal DOFs with the COM motion. Firstly, we demonstrate, for the first time to our knowledge, levitation in high vacuum of a nanodiamond hosting a single NV center. By trapping with a Paul trap, we avoid particle photo-damage, and with low optical intensity are able to measure the NV single photon emission and monitor the nanodiamond's COM motion (see Figure \ref{fig:paultrap}). Secondly, we implement a feedback scheme to reduce the effective temperature of this COM motion and improve the particle's stability and confinement at low pressures. 

% get a bit more specific on how we do it, before the methods section

%To this aims, we designed an end-cap Paul trap configuration, with high optical access, that allows us for the first time to detect and characterize single photons from nanoparticles trapped in a Paul trap. Additionally, by relying on electric fields to levitate particles instead of optical fields, we directly avoid particle heating, which is a significant advantage over optical tweezers experiments. The powers required to continuously excite the NV transition are two orders of magnitude lower than those used for optical trapping\citep{Neukirch2015}. This lets us work at pressures of the order of $10^{-2}$ mBar. [THIS PART IS TOO SPECIFIC TO BE LOCATED IN THIS PARAGRAPH]

%%%%%%%%%%%%%%%%%%%%%%%%%%%%%%%%%%%%%%%%%%%%%%%%%%%%%%%%%%%%%%%%%%%%%
%%\section{Methods // Experimental setup} 3rd
%%%%%%%%%%%%%%%%%%%%%%%%%%%%%%%%%%%%%%%%%%%%%%%%%%%%%%%%%%%%%%%%%%%%%

% Describe Paul trap: just a couple of equations, some intuition, brownian motion, technical details

The COM equation of motion can be found by calculating the force experienced by a charged particle in the Paul trap time-dependent electric potential\citep{Paul1990}:
\begin{align}
\Phi = \frac{V_0 \cos \omega t}{2 d^2}(2z^2 - r^2).
\end{align}
where $V_0$ is the voltage amplitude, $\omega$ the driving frequency, $t$ the time variable, $d$ a geometric constant with length units and $z$, $r$ are cylindrical coordinates. Even though the particle's COM Hamiltonian $H(t)$ is explicitly time-dependent, under certain general conditions the motion can be averaged over the high frequency $\omega$\citep{Major2006}. This is known as the adiabatic approximation and ensures that the particle is governed by an effective potential $U_\text{eff} = \frac{1}{2}\Omega_i^2 x_i^2$ for every axis $x_i$, where $\Omega_i$ is known as the secular frequency of the $i$-th axis.

Therefore, for low frequencies the full equation of motion for the particle's COM along the $z$ axis can be simplified in the following way
\begin{align}
\mathrm{d}p_z + \gamma \mathrm{d}z  - \frac{2qV_0}{d^2}\cos (\omega t)z\mathrm{d}t = \sigma \mathrm{d}W_t \quad \rightarrow \quad \mathrm{d}p_z + \gamma  \mathrm{d}z + k_\text{eff} z = \sigma \mathrm{d}W_t \, ,
\end{align}
where $p_z$ is the momentum in the $z$ direction, $\gamma$ the damping constant, $q$ the charge of the particle, $k_\text{eff} = m\Omega_i^2$ the restoring force of the effective potential and $\sigma \mathrm{d}W_t$ a stochastic force with standard deviation $\sigma$, associated with the damping via the fluctuation-dissipation relation $\sigma = \sqrt{2 k_B T \gamma}$, where $k_B$ is Boltzmann's constant and $T$ is temperature\citep{Kubo1966}. By reducing the pressure and hence $\gamma$, the motion decoherence is reduced, while the inversely proportional $Q$ factor of the oscillator increases. This is why a trapping scheme capable of bringing particles to low pressures is appealing.

\begin{figure}
\includegraphics[width=0.8\textwidth]{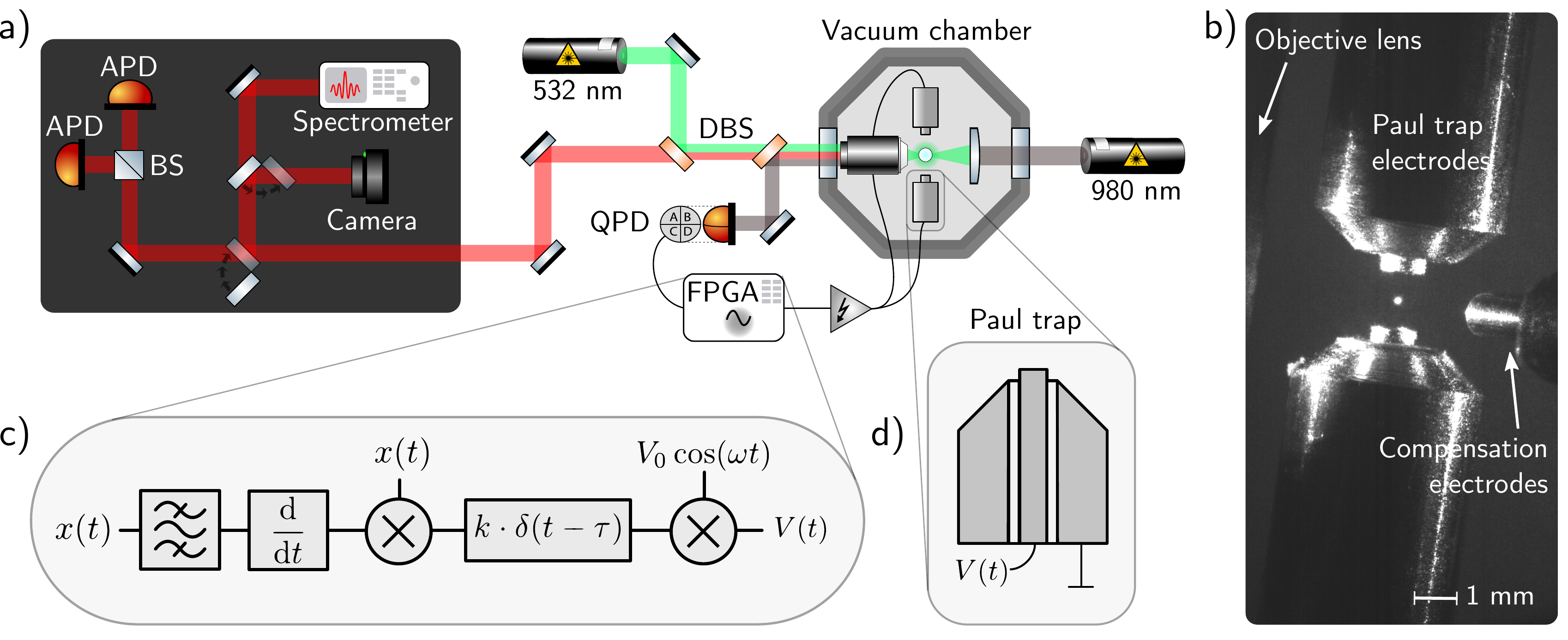}
\caption{a) Sketch of the experimental setup. A 532 nm CW laser excites the NV center. The emitted fluorescence is collected back through the same microscope objective (Olympus long working distance, NA 0.8), filtered, and sent to a light-proof box for fluorescence detection. The particle's motion is monitored by focusing a 980 nm laser onto the particle and detecting the forward scattered light with a QPD. 
b) Camera image of the trap viewed from above, showing the end-cap Paul trap electrodes, compensation electrodes, and the objective lens. 
c) Signal processing blocks of the feedback system. The particle's motion $x(t)$ is digitally acquired in the FPGA 
and band-pass filtered to eliminate noise and unwanted modes. Then, its derivative $\dot{x}(t)$ is numerically calculated and the product $x(t)\dot{x}(t)$ is used to modulate the amplitude of the driving RF signal. The modulation amplitude and delay are set to values that minimize the mode energy. 
d) Cutaway of a Paul trap electrode. Particles are levitated at the trap's center of symmetry, above the inner rod. The trapping pseudo-potential is generated by grounding the outer electrodes and applying the driving radio-frequency signal $V(t)$ to the inner ones; a white alumina tube is used for electrode insulation.}
\label{fig:setup}
\end{figure}

Our trap design (see Figure \ref{fig:setup}b) has an endcap geometry for good optical access. It is made of two assembled steel electrodes separated by 1.4 mm mounted on a ceramic holder. These electrodes are mounted on a three axis piezoelectric stage (Figure \ref{fig:setup}a) and driven by a high voltage signal generated by a field programmable gate array (FPGA) card and a high voltage amplifier (Matsusada AMT-1B60). The FPGA acts as a wave generator, and provides a sinusoidal output at adjustable frequency and amplitude. Usual working parameters in our experiment are driving frequency $\omega/2\pi = 20$ kHz and amplitude $V_0$ in the range of 0.75 kV$_\text{pp}$ to 2 kV$_\text{pp}$. Two extra rods pointing towards the center of the trap act as compensation electrodes. The latter are used to eliminate stray fields and thus minimize residual micromotion driving.

% loading mechanism

To load particles into the Paul trap, we use electrospray injection at ambient pressure with a suspension of nanodiamonds in ethanol (Adámas, 40 nm diameter nanodiamonds, 1-4 NV; see Supplementary information). Trapping events are monitored with a camera (Thorlabs CMOS), using the scattering from a weakly focused 980 nm laser (Figure \ref{fig:setup}b). The Paul trap is subsequently moved with the piezoelectric stages to bring the particle into the focus of a high numerical aperture (NA) microscope objective (Olympus LMPLFLN 100x). The objective is used to illuminate the nanodiamond with a 532 nm laser, and thus excite the NV transition, and also to collect the emitted fluorescence. The fluorescence light is then directed towards a light-proof box, where it is selectively analysed with an electron multiplying camera (EMCCD) (Andor iXon+ EMCCD), a spectrometer (Acton Spectrapro 2500i), or coupled into low dark-counts single photon detectors (Perkin Elmer SPCM-AQR-14). Due to the low intensity of the two laser beams, we did not detect any effect of the dipole force on the particle dynamics in the Paul trap. 

% optical setup and fluorescence collection

The single photon detectors are set up in a Hanbury, Brown and Twiss (HBT) configuration to study the emitters photon statistics\citep{Brown1956}. They are used to measure the time dependent intensity correlation function, defined as 
\begin{align}
g^{(2)}(\tau) = \frac{\langle : I(t)I(t + \tau) :\rangle}{\langle I(t) \rangle^2}
\end{align}
where $I$ is the intensity operator and ${\langle : \ldots :\rangle}$ indicates normal ordering. For a single photon emitter,  $g^{(2)}(0) = 0$ and for two or more (equal) emitters $g^{(2)}(0) \geq 0.5$. Usually, background contributions lead to a $g^{(2)}(0) > 0$ even though the main contribution stems from a single emitter, which is ensured by $g^{(2)}(0) < 0.5$\citep{Gardiner2004}. This condition is the criterion we use to identify trapped nanodiamonds holding single NVs. 

% Pumping down

After trapping at ambient pressure we characterize the fluorescence emission of the loaded nanodiamonds. Fluorescence imaging of the particles is performed with the EMCCD camera, followed by $g^{(2)}(\tau)$ measurements. We have detected single NVs in trapped nanodiamonds, with usual count rates of 2000 to 7000 counts per second at excitation powers of 2 - 5 mW. Even though the rotation of the trapped nanodiamonds has not been studied in this experiment, measuring and controlling the crystal Euler angles will be important for coupling internal and external degrees of freedom, as the spin interaction depends on the NV orientation. This could be implemented by using elongated nanodiamonds, whose birefringence will transduce the particle's rotation into a modulation of the intensity and a change in the light polarization\citep{Simpson2007}\citep{Arita2013}\citep{Schell2017flying}.

After a single NV was detected, we compensated the stray fields to minimize the micromotion driving. Stray field compensation was repeated for every new particle, since it was prone to drifts and sign changes. This may be caused by the introduction of other charged particles in the chamber during the use of the electrospray, which can change the static electric field in the trapping region. By following the described procedure, particles can be stably trapped at ambient pressure over very long times: some levitated nanodiamonds were kept for weeks, until they were deliberately substituted by other nanocrystals. 

At ambient pressure, the stochastic forces due to collisions with air molecules damp oscillations and lead to overdamped Brownian dynamics. Thus, when a trapped single NV is detected, we decrease the pressure to bring the chamber to vacuum. The voltage amplitude used to trap particles is 2 kV$_\text{pp}$, but Paschen's law predicts a minimum of the breakdown voltage of air at 750 V$_\text{pp}$ \citep{Paschen1889}. Voltages above this value will ignite a plasma at pressures that depend on the electrodes geometry, leading to particle loss. Consequently, while decreasing the pressure we reduce the voltage amplitude below 750 V$_\text{pp}$; at pressures lower than $5\cdot 10^{-1}$ mBar the voltage amplitude can be safely increased again.

% feedback system

For pressures below 1 mBar, the motion coherence of the particle is sufficient to activate the feedback and reduce the energy of the particle. The feedback system, described in Figure \ref{fig:setup}c, is implemented in the following way: a 980 nm laser is focused with a low effective NA and superimposed with the focus of the trap's center. The focused light together with the light scattered by the particle are collected by the high NA objective. The 980 nm light is then directed onto a quadrant photo-diode (QPD), which extracts signals that are proportional to the particle's position $V_i(t) \propto x_i(t)$ for the $i$-th axis. For the $x$ and $y$ axis, this signal is obtained via differential measurements of the intensity on the quadrants. Conversely, the dynamics in $z$ are detected from a modulation on the total intensity, resulting from the interference between the laser beam and the position dependent particle scattering. These signals are fed to the FPGA and processed (Figure \ref{fig:setup}b). The FPGA modulates the amplitude of the driving voltage to cool down the energy of the COM motion $i$\citep{Nagornykh2015}. The modulation is obtained by calculating the product $k\cdot V_i(t) \dot{V}_i(t)$, which is effectively at a frequency of $2\Omega_i$. Here, $k$ is an experimentally determined constant that can be different for different particles.

%%%%%%%%%%%%%%%%%%%%%%%%%%%%%%%%%%%%%%%%%%%%%%%%%%%%%%%%%%%%%%%%%%%%%
%%\section{Results and discussion} 2nd 
%%%%%%%%%%%%%%%%%%%%%%%%%%%%%%%%%%%%%%%%%%%%%%%%%%%%%%%%%%%%%%%%%%%%%

% how it works (the trap), usual frequencies at vacuum, number of charges, position readout 

\begin{figure}
\includegraphics[width=0.8\textwidth]{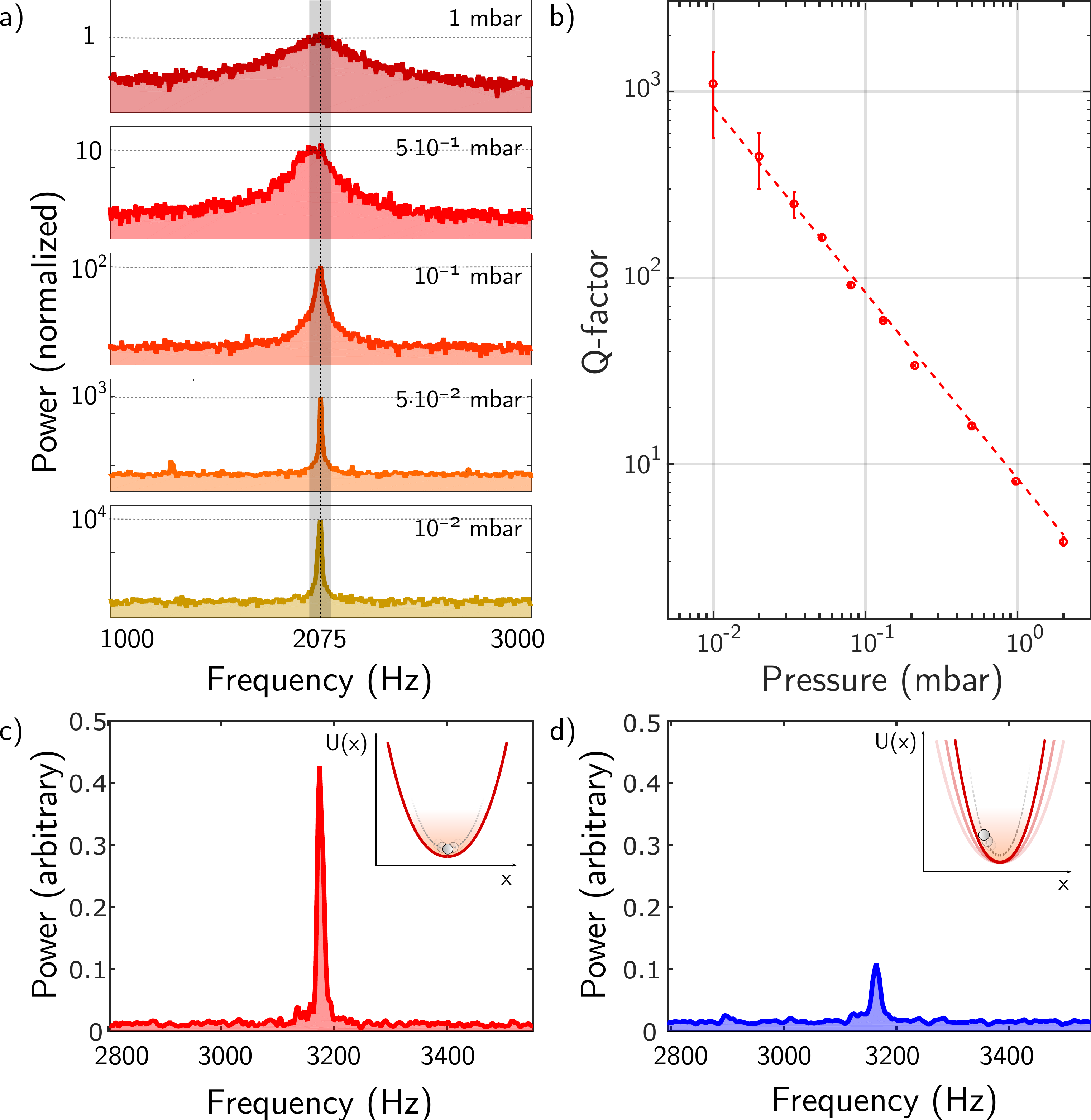}
\caption{Measurements of the particle's motion and feedback. The translational modes are driven by thermal noise, which is dominant for Brownian particles such as 40 nm nanodiamonds. 
a) Power spectral density in log-scale of the particle's COM, showing increasingly better defined resonance peaks at lower pressures. Approximate values of the powers at the peaks are normalized to the first plot maximum; the area under the curves is proportional to the particle's COM temperature. The flat part of the spectrum is dominated by laser shot noise. 
b) $Q$ factor inferred by fitting the measured power spectral densities of a single nanodiamond at different pressures to the one expected from a harmonic oscillator driven by Brownian noise (see Supplementary information). The bigger error bars at low pressures are due to fitting uncertainties, since most of the oscillator response is buried in the noise floor. 
c) and d): Temperature reduction of the energy in the eigenmode by a factor of 4, corresponding to a temperature of 75 K. In panel c) the feedback is off (no potential modulation) while it is on in panel d) (potential modulated at $2\Omega$). The measurements are performed on a 40 nm diamond at $5\cdot 10^{-1}$ mBar at the same time as a $g^{(2)}(\tau)$ intensity correlation measurement was being taken (see Figure \ref{fig:singleNV}).}
\label{fig:signal}
\end{figure}

Particles at low pressures were much more stable when stray fields had been cancelled with the compensation electrodes: without 532 nm excitation light and only some weak 980 nm illumination to detect the COM, we were able to trap nanodiamonds at pressures as low as $10^{-3}$ mBar, where the particle's oscillating motion attained $Q$ factors above $10^3$ (Figure \ref{fig:signal}b); at lower pressures the nanodiamonds generally became unstable. The cancellation of the stray fields was also required for feedback cooling. With it, the energy of the axial COM mode could be reduced to a fraction of its value at thermal equilibrium $\Bigl\langle  m \Omega^2 {x}^2 \Bigr\rangle= k_B T$ expected by the equipartition theorem. Figure \ref{fig:signal} shows the effect of the feedback on a nanodiamond levitated at $5\cdot 10^{-1}$ mBar, demonstrating a reduction of energy of the eigenmode. We achieved energy reductions of 6 to 9 dB at pressures between $10^{-2}$ mBar and $10^{-1}$ mBar, which corresponds to effective temperatures in the range of 75 K to 38 K. This reduction in energy is partly limited by the signal to noise ratio (SNR) from the signal measured using the QPD, but the biggest restriction is the level of vacuum. Indeed, at lower pressures the particle's motion is more predictable and the feedback performance improves. 

The pressure level is, however, limited by diamond photo-degradation. Although nanodiamonds in Paul traps do not inherently suffer from heating due to optical absorption, a certain amount of optical power is required to interact with the single NV center, whether for excitation or for motion detection. As a result, laser power must be high enough for a good SNR, yet sufficiently low to avoid heating up the particle excessively. At ambient pressure, this limitation is not present since the interaction with air molecules not only damps oscillations but is also sufficiently high to keep the particle's lattice temperature at equilibrium with the environment. In the future this constraint can be eliminated, since spin initialization and readout only require short laser pulses, hence allowing us to work at lower pressures.

% Feedback system: pressures and performance say something about particle shrinking

In order to probe the capacity of the nanodiamond to withstand heat dissipation, we performed $g^{(2)}(\tau)$ measurements at different pressures. We obtained stable fluorescence counts up until the $10^{-1}$ mBar to $5\cdot 10^{-1}$ mBar range, with 1 to 3 mW of 532 nm power; no signature of quenching due to high temperatures was observed under these conditions\citep{Toyli2012}. At this point, a progressive decrease in the scattering signal is apparent, indicating a possible shrinking of the nanodiamonds. 

% Fluorescence: g2 correlation measurements, spectrum with ethanol Raman

\begin{figure}
\includegraphics[width=0.8\textwidth]{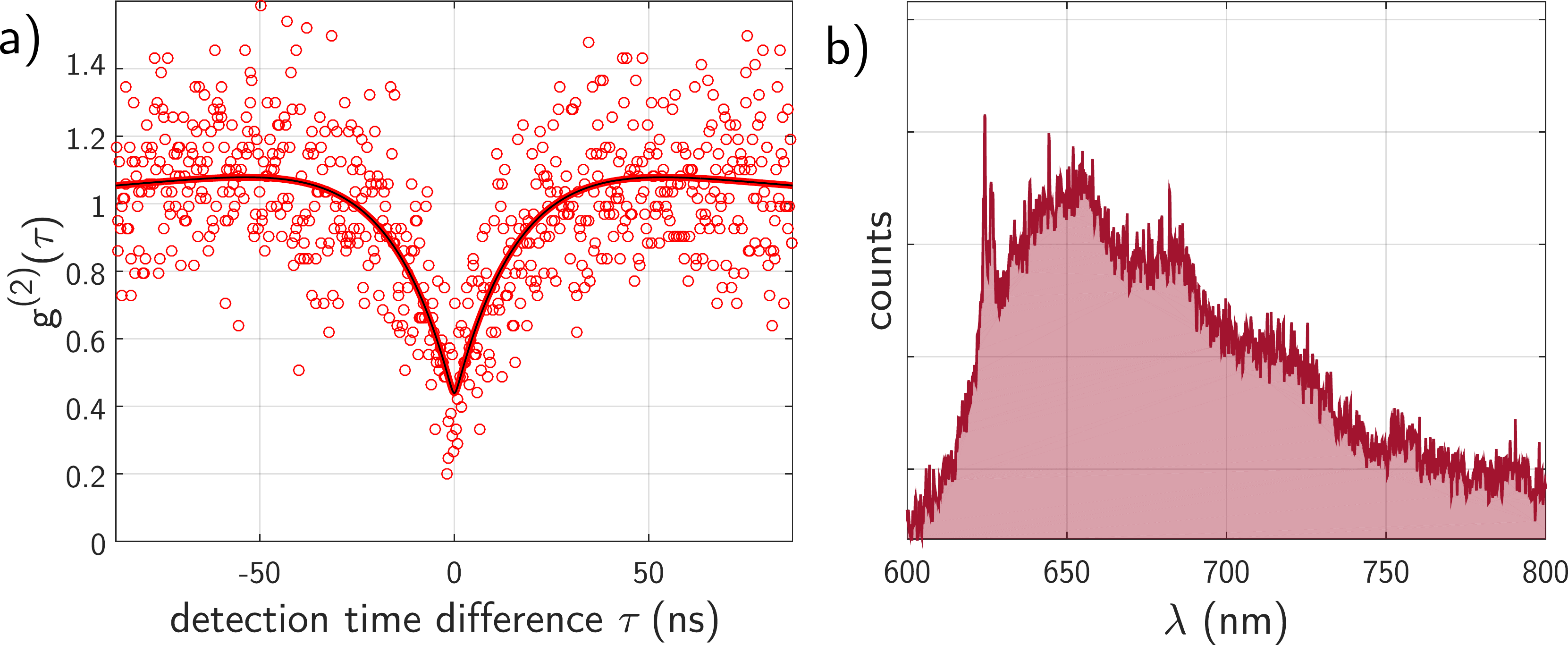}
\caption{
a) $g^{(2)}(\tau)$ intensity correlation measurement of the NV$^-$ fluorescence of 
a nanodiamond with applied feedback (data shown in Figure \ref{fig:signal}). 
The correlation measurement is fitted with a modified three-level model\citep{Kurtsiefer2000} convoluted with the instrument response function, with $g^{(2)}(\tau)$ showing a minimum at $0.4 \pm 0.1$, and thus the presence of a single NV center. Other $g^{(2)}(\tau)$ measurements for different nanodiamonds are shown in the Supplementary information. 
b) Fluorescence emission spectrum of the single NV center. The two peaks at 625 nm are due to Stokes Raman scattering, and indicate the presence of ethanol traces from the electrospray in the particle.}
\label{fig:singleNV}
\end{figure}

Figure \ref{fig:singleNV}a shows the $g^{(2)}(\tau)$ measurement and spectrum of a single NV in a nanodiamond trapped at 0.5 mBar, the same on which we applied the feedback shown in Figure \ref{fig:signal}d. Figure \ref{fig:singleNV}b shows the fluorescence emission spectrum obtained for this NV. The two peaks at $\lambda = 620$ nm are caused by Raman scattering (OH stretching) of ethanol traces\citep{Dolenko2015} present in the nanodiamonds.

%%%%%%%%%%%%%%%%%%%%%%%%%%%%%%%%%%%%%%%%%%%%%%%%%%%%%%%%%%%%%%%%%%%%%
%%\section{Conclusions} 5th
%%%%%%%%%%%%%%%%%%%%%%%%%%%%%%%%%%%%%%%%%%%%%%%%%%%%%%%%%%%%%%%%%%%%%

In conclusion, we have demonstrated stable trapping of nanodiamonds and optical readout of a single NV center at pressures down to $5\cdot 10^{-1}$ mBar. This is two orders of magnitude lower than previously attained with optical traps\citep{Neukirch2015}, where thermal damage limits access to higher vacuum levels. Furthermore, our platform enables us to cool down the particle's translational modes via parametric feedback. In future experiments, control over the spin system of the levitated NV center can be attained by designing and adding a microwave antenna in the trap. Having such a high level of control of a particle hosting emitters is a decisive step towards achieving strong coupling between a spin and the particle's mechanical degrees of freedom\citep{Rabl2009}\citep{Yue2017}\citep{Delord2017strong}.

%The blocks of the forthcoming quantum systems and networks will require fine control of its mechanical components, 
%and with hybrid systems it will be possible to address them indirectly by acting on coupled quantum degrees of freedom. 
%This work paves the way towards the construction of quantum systems based on the coupling of internal and external degrees of freedom in levitated nanoparticles.

%%%%%%%%%%%%%%%%%%%%%%%%%%%%%%%%%%%%%%%%%%%%%%%%%%%%%%%%%%%%%%%%%%%%%
% ASSOCIATED CONTENT
%%%%%%%%%%%%%%%%%%%%%%%%%%%%%%%%%%%%%%%%%%%%%%%%%%%%%%%%%%%%%%%%%%%%%
\section{Associated content}

The supplementary material contains information regarding the particle dispersion, loading of particles to the Paul trap, oscillation frequencies and $Q$ factor, excess charge determination, feedback system and $g^{(2)}(\tau)$ measurements.

%%%%%%%%%%%%%%%%%%%%%%%%%%%%%%%%%%%%%%%%%%%%%%%%%%%%%%%%%%%%%%%%%%%%%
% ACKNOWLEDGEMENTS
%%%%%%%%%%%%%%%%%%%%%%%%%%%%%%%%%%%%%%%%%%%%%%%%%%%%%%%%%%%%%%%%%%%%%

\begin{acknowledgement}
The authors acknowledge financial support from the European Research Council through grant QnanoMECA (CoG - 64790), Fundació Privada Cellex, CERCA Programme / Generalitat de Catalunya, and the Spanish Ministry of Economy and Competitiveness through the Severo Ochoa Programme for Centres of Excellence in R$\&$D (SEV-2015-0522), grant FIS2016-80293-R, and Juan de la Cierva grant IJCI-2015-26091.
\end{acknowledgement}

%%%%%%%%%%%%%%%%%%%%%%%%%%%%%%%%%%%%%%%%%%%%%%%%%%%%%%%%%%%%%%%%%%%%%
%% BIBLIOGRAPHY
%%%%%%%%%%%%%%%%%%%%%%%%%%%%%%%%%%%%%%%%%%%%%%%%%%%%%%%%%%%%%%%%%%%%%

\bibliography{./references}

\newpage
\section{Supplementary information}
\subsection{Particle dispersion}
Nanodiamonds from Adámas Nanotechnologies (40 nm in diameter, 1-4 NV) were used for the experiment. 20 $\mu$l of the commercial dispersion were dispersed in 40 ml of ethanol and pumped to the electrospray using a syringe. To avoid particle clustering, we utilised a syringe filter with 100 nm pore size. Furthermore the dispersion was sonicated for 10 minutes before the experiment.

\subsection{Particle loading}
A custom made electrospray system was introduced into the chamber at ambient pressure, and a camera and a 980 nm laser were used to display and illuminate the trapping region in real time. Particles were injected with the electrospray by activating a 3 kV voltage source and then pumping the dispersion. Particles ejected with the electrospray would pass through the center of the trap potential. Some would lose enough energy due to air damping to stay trapped in the potential, and it was usual to trap several at once. Excess particles were pushed out of the trap by carefully reducing the voltage. After only a single particle remained, the electrospray was removed and the vacuum chamber closed. 

\subsection{Oscillation frequencies and $Q$-factor calculation}

Under the adiabatic approximation, the power spectral density of the particle's COM will take the expression
\begin{align}\label{PSD}\tag{S1}
S_x(\Omega) = \frac{\sigma^2/m^2}{\left(\frac{\Omega_i}{Q}\right)^2\Omega^2 + \left(\Omega_i^2  - \Omega^2\right)^2},
\end{align}
where $m$ is the particle's mass, $\sigma = \sqrt{2k_BT\gamma}$ is the Brownian noise standard deviation, $\Omega_i$ the secular angular frequency of the oscillator and $Q$ its quality factor. By introducing the reduced damping constant $\Gamma = \frac{\gamma}{m}$, which for low damping corresponds to the full width at half maximum of the oscillator response, the quality factor is defined as $Q = \frac{\Omega_i}{\Gamma}$. By fitting the measured power spectral density to expression \ref{PSD}, we can indirectly measure the value of $Q$ and $\Omega_i$.

Observed secular frequencies in the trap axis direction (for $V_{\text{pp}} = 750$) are specified in table \ref{table:secular_frequencies}. 

\begin{table}[h!]
\centering
\begin{tabular}{||c c c||} 
 \hline
 Min. freq & Typical & Max. freq\\
 \hline\hline
-- &  (1.5 kHz, 6 kHz) & 15 kHZ \\ 
 \hline
\end{tabular}
\caption{Secular oscillation frequencies, observed in trapped nanodiamonds}
\label{table:secular_frequencies}
\end{table}

\subsection{Excess charge determination}

The charge of the particle can be calculated as follows: since the particle's mass is known (this is a reasonable assumption, given the fact that we filter the particles by size with a porous filter). In the underdamped regime (i.e., low damping by air molecules), the equation of motion for the center of mass of the particle is described by the Mathieu\citep{Major2006} equation,
\begin{align}\label{eq:parameter_q}\tag{S2}
\ddot{u}(\tau) - 2q \cdot \cos (2\tau) u(\tau) = 0,
\end{align}
where $q = \frac{2\mathcal{Q}V}{md^2\omega^2}$. Here, $\mathcal{Q}$ is the particle charge (not to be confused with $Q$, the quality factor), $V$ the voltage, $m$ the particle mass, $d$ a geometrical parameter (which is constant and can be calculated for a given trap) and $\omega$ the driving angular frequency. Since we set the voltage and frequency, the only unknown parameter is $\mathcal{Q}$. Thus, if the parameter $q$ can be measured, we can solve equation \ref{eq:parameter_q} for $\mathcal{Q}$ and find the particle charge. 

This parameter $q$ can be found indirectly. Firstly, we measure the ratio between the particle secular frequency $\Omega$ and the driving frequency, $\beta = 2\frac{\Omega}{\omega}$. $\beta$ is a characteristic exponent of the Mathieu's equation, and is a function of $q$ only: $\beta(q)$. Therefore, by numerically inverting the expression $\beta(q)$ we get $q(\beta)$. Finally, by plugging the obtained value and calculating 
\begin{align}\tag{S3}
\mathcal{Q} = \frac{q(\beta)md^2\omega^2}{2V}
\end{align}
we can estimate the particle charge. Usual elementary charge numbers that we have measured this way lay in the range of 40 to 150 e$^+$.

\subsection{Feedback}
The radio frequency signal applied to the inner rod has the expression 
\begin{align}\tag{S4}
V(t) = V_0(1 + k\cdot V_i(t)\dot{V}_i(t))\cos(\omega t).
\end{align}
The spectrum of the particle motion $x_i(t)$ has a peak at $\Omega_i$, corresponding to its secular frequency. Since the feedback is a parametric modulation, $V(t)$ will enter the equation of motion as a product with $x(t)$. By a trigonometric identity, this product results in two driving signals at $\Omega_i$ and $3\Omega_i$ components, but since the $3\Omega_i$ is off-resonance, this component can be neglected. Thus, the modulation will have approximately a driving effect on the motion of the particle: by adjusting the delay accordingly (in phase or at push-pull), we can make it excite or damp the oscillation. 

If all three axis need to be cooled, the secular frequencies of the trap requires different values for every axis; otherwise, cooling one axis could result in heating another with the same degenerate frequency, since phase between translational modes is uncoupled. Although our trap has symmetry of revolution, and thus two equal eigenfrequencies, this symmetry can be broken to separate all three frequencies by slightly modifying the electrodes. 

\subsubsection{Effective temperature}
When the nanoparticle is close to the potential minimum of a deep trap, the oscillations can be treated as approximately harmonic. In that case, the average number of vibrational excitations $\bar{n}$ at thermal equilibrium is given by the Bose-Einstein distribution
\begin{align}\tag{S5}
\frac{\bar{E}}{\hbar \Omega} = \bar{n}_\mathrm{eq}  = \frac{1}{2} + \left ( e^{\hbar \Omega/k_B T} - 1 \right)^{-1},
\end{align}
where $\Omega/2\pi$ is the frequency of harmonic oscillations. In the limit of high temperature this reduces to the equipartition expression (via first order Taylor expansion)
\begin{align}\tag{S6}
\bar{E} = \hbar\Omega\bar{n}_\mathrm{eq} \approx k_B T,
\end{align}
where $\bar{E}$ is the average energy. Therefore, given some experimentally determined $\bar{n}_\mathrm{eq}$ or $\bar{E}$, one can obtain an effective temperature according to the above formulae, whether or not the system is truly at equilibrium.

\subsection{Intensity correlation measurements}
An Olympus LMPLFLN 100x objective, with NA 0.8, was used to collect and collimate the fluorescence of the NV center, and single photons were detected with Perkin Elmer SPCM-AQR-14 APDs. Single photon counting was performed with a PicoHarp 300 system. 

\subsubsection{Background correction}
For continuous excitation, the joint second order correlation function of an emitter with intensity $\langle n_s \rangle = s$ and a background noise with intensity $\langle n_b \rangle = b$ will be
\begin{align}\tag{S7}
g^{(2)}_{sb} &= \frac{\langle \left(s(t) + b(t)\right)\left (s(t + \tau) + b(t + \tau)\right)\rangle}{(s + b)^2}\\
g^{(2)}_{sb} &= \frac{s^2 g^{(2)}_s + b^2 + 2sb}{(s + b)^2},\tag{S8}
\end{align}
where $g_s^{(2)}$ is the emitter second order correlation function and it has been assumed that
\begin{itemize}
\item The background noise is quantum mechanically uncorrelated with the emitter fluorescence
\item $g_b^{(2)} = 1$
\end{itemize}
Therefore, by measuring $s$ and $b$ it is possible to correct $g^{(2)}_{sb}$ (which is the function measured in our single photon counting system) to obtain the emitter $g^{(2)}_{s}$.

Analogously, by defining $\rho = \frac{s}{s + b}$, we obtain
\begin{align}\tag{S9}
g^{(2)}_s = \frac{g^{(2)}_{sb} + \rho^2 - 1}{\rho^2}.
\end{align}

Background noise was measured to be $150 + f(I)$ counts per second, where 150 was the average dark count rate in our APDs and $f(I)$ was a pumping laser intensity dependent fluorescence contribution, due to optical elements in our setup. 
\subsubsection{Model fitting}
\begin{figure}\renewcommand\figurename{Figure S11}
  \includegraphics[width=0.8\textwidth]{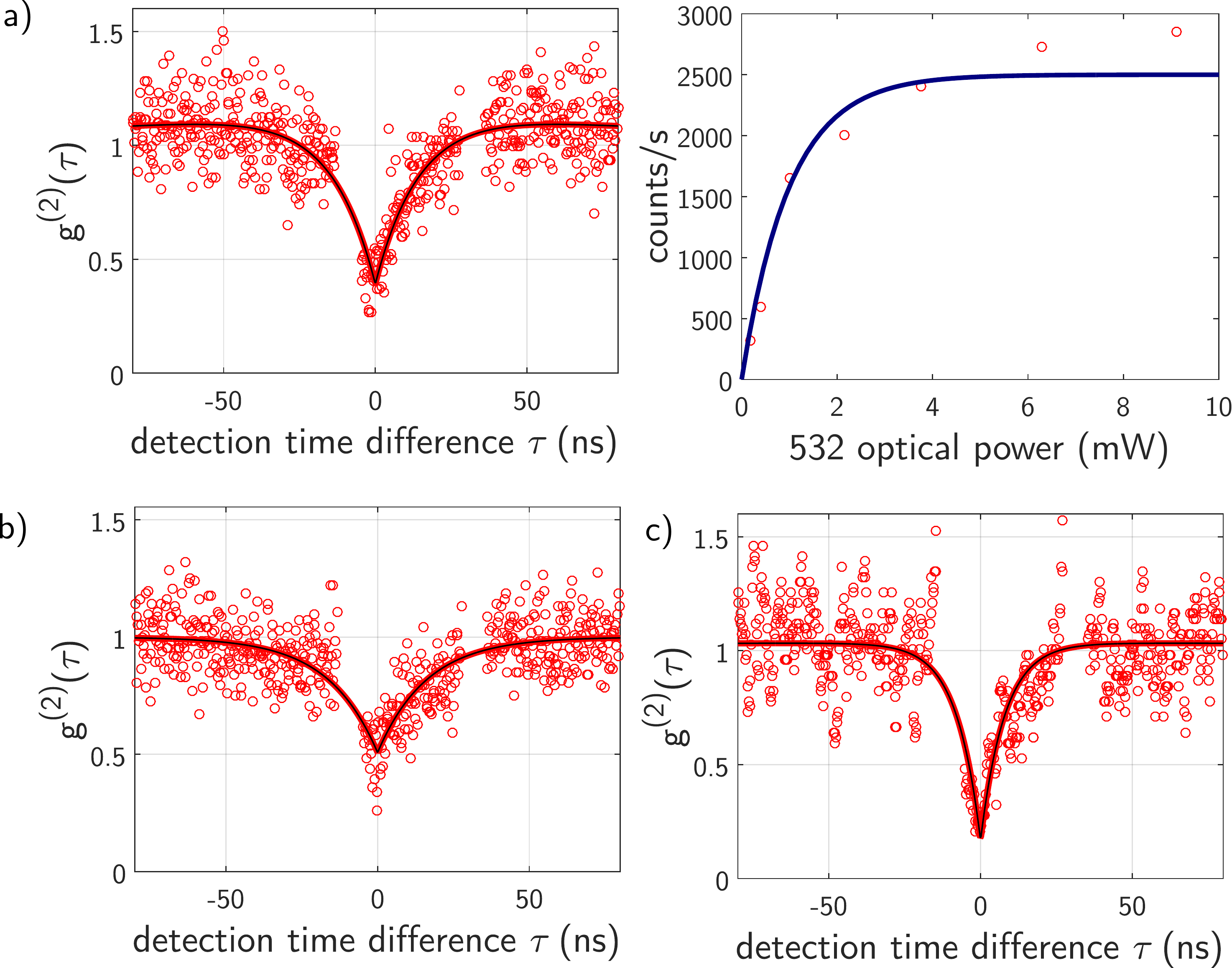}
  \caption{a) Single NV $g^{(2)}(\tau)$ measurement along with saturation curve. Two sets of data, one at each side of the dip, were deleted, because when the APDs detected a photon they emitted new fluorescence, which the other APD would falsely detect as coming from the emitter (PicoQuant SAPD). After the PicoQuant APD was replaced by a Perkin Elmer, this stopped being a problem. b) and c) show further measurements of a 2NV and a single NV nanodiamond}
  \label{fig:other_particles}
\end{figure}
The sets of measured $g^{(2)}(\tau)$ functions are fitted with a least squares criterion to a modified three-level model\citep{Kurtsiefer2000} after background subtraction
\begin{align}\tag{S10}
g^{(2)}(\tau) = 1 + p^2_f\left(c e^{-\frac{|\tau|}{\tau_1}} - (1 + c)e^{-\frac{|\tau|}{\tau_2}}\right).
\end{align}
Since we do not measure $g^{(2)}(\tau)$ but $g^{(2)}(\tau) * h(\tau)$, where $h(\tau)$ is the APD system instrument response function, we fit the data with $g^{(2)}(\tau) * h(\tau)$ instead. With the obtained parameters, we calculate $g^{(2)}(0) = 1 - p_f^2$ to determine whether the emitter is a single photon source or not. The confidence levels for $g^{(2)}(0)$ with $\pm \sigma$ where calculated from the inferred $p_f$ distribution.

Measurements of three particles are shown in Figure \hyperref[fig:other_particles]{S11}.

\end{document}